\documentclass[%
  aip,
  amsmath,amssymb,
  reprint,%
]{revtex4-1}

\usepackage{graphicx} % Include figure files
\usepackage{dcolumn}  % Align table columns on decimal point
\usepackage{bm}       % bold math

\usepackage[utf8]{inputenc}
\usepackage[T1]{fontenc}

\usepackage{mathptmx}
\usepackage{amsfonts}
\usepackage{mathrsfs}

\usepackage{booktabs}
\usepackage{makecell}
\usepackage{siunitx}

\usepackage{etoolbox}
\usepackage{soul}
\usepackage{wrapfig}
\usepackage{xcolor}
\usepackage{float}
\usepackage{algorithm}
\usepackage{algorithmic}
\usepackage[hidelinks]{hyperref}
\usepackage{lineno}

\makeatletter
\def\@email#1#2{%
 \endgroup
 \patchcmd{\titleblock@produce}
  {\frontmatter@RRAPformat}
  {\frontmatter@RRAPformat{\produce@RRAP{*#1\href{mailto:#2}{#2}}}\frontmatter@RRAPformat}
  {}{}
}%

\makeatletter
\def\switch@array{}
\makeatother

\begin{document}

\preprint{AIP/123-QED}

\title[]{Reinforcement learning with reputation-based adaptive exploration promotes cooperation}
% Force line breaks with \\

% 1
\affiliation{School of Mathematical Sciences, Beihang University, Beijing 100191, China}

% 2
\affiliation{School of Artificial Intelligence, Beihang University, Beijing 100191, China}

% 3
\affiliation{Hangzhou International Innovation Institute, Beihang University, Hangzhou 311115, China}

% 4
\affiliation{Institute of Trustworthy Artificial Intelligence, Zhejiang Normal University, Hangzhou 310013, China}

% 5
\affiliation{Key laboratory of Mathematics, Informatics and Behavioral Semantics, Beihang University, Beijing 100191, China}

% 6
\affiliation{Zhongguancun Laboratory, Beijing 100094, China}

% 7
\affiliation{Institute of Medical Artificial Intelligence, Binzhou Medical University, Yantai 264003, China}

% 8
\affiliation{Beijing Advanced Innovation Center for Future Blockchain and Privacy Computing, Beihang University, Beijing 100191, China}

% 9
\affiliation{Beijing Academy of Blockchain and Edge Computing, Beijing 100085, China}

% 10
\affiliation{State Key Laboratory of General Artificial Intelligence, BIGAI, Beijing 100080, China}

% 11
\affiliation{School of Mathematics and Statistics, Nanjing University of Science and Technology, Nanjing 210094, China}

\author{An Li}
\thanks{These authors contributed equally to this work.}
\affiliation{School of Mathematical Sciences, Beihang University, Beijing 100191, China}
\affiliation{Key laboratory of Mathematics, Informatics and Behavioral Semantics, Beihang University, Beijing 100191, China}

\author{Wenqiang Zhu}
\thanks{These authors contributed equally to this work.}
\affiliation{School of Artificial Intelligence, Beihang University,  Beijing 100191, China}
\affiliation{Key laboratory of Mathematics, Informatics and Behavioral Semantics, Beihang University, Beijing 100191, China}
\affiliation{Zhongguancun Laboratory, Beijing 100094, China}

\author{Chaoqian Wang}
\affiliation{School of Mathematics and Statistics, Nanjing University of Science and Technology, Nanjing 210094, China}

\author{Longzhao Liu}
\affiliation{School of Artificial Intelligence, Beihang University, Beijing 100191, China}
\affiliation{Key laboratory of Mathematics, Informatics and Behavioral Semantics, Beihang University, Beijing 100191, China}
\affiliation{Zhongguancun Laboratory, Beijing 100094, China}
\affiliation{Beijing Advanced Innovation Center for Future Blockchain and Privacy Computing, Beihang University, Beijing 100191, China}

\author{Hongwei Zheng}
\affiliation{Beijing Academy of Blockchain and Edge Computing, Beijing 100085, China}

\author{Yishen Jiang}
\affiliation{School of Mathematical Sciences, Beihang University, Beijing 100191, China}
\affiliation{Key laboratory of Mathematics, Informatics and Behavioral Semantics, Beihang University, Beijing 100191, China}
\affiliation{Zhongguancun Laboratory, Beijing 100094, China}

\author{Xin Wang}
\email{wangxin\_1993@buaa.edu.cn}
\affiliation{School of Artificial Intelligence, Beihang University, Beijing 100191, China}
\affiliation{Key laboratory of Mathematics, Informatics and Behavioral Semantics, Beihang University, Beijing 100191, China}
\affiliation{Zhongguancun Laboratory, Beijing 100094, China}
\affiliation{Beijing Advanced Innovation Center for Future Blockchain and Privacy Computing, Beihang University, Beijing 100191, China}
\affiliation{State Key Laboratory of General Artificial Intelligence, BIGAI, Beijing 100080, China}

\author{Shaoting Tang}
\email{tangshaoting@buaa.edu.cn}
\affiliation{School of Artificial Intelligence, Beihang University, Beijing 100191, China}
\affiliation{Hangzhou International Innovation Institute, Beihang University, Hangzhou 311115, China}
\affiliation{Institute of Trustworthy Artificial Intelligence, Zhejiang Normal University, Hangzhou 310013, China}
\affiliation{Key laboratory of Mathematics, Informatics and Behavioral Semantics, Beihang University, Beijing 100191, China}
\affiliation{Zhongguancun Laboratory, Beijing 100094, China}
\affiliation{Institute of Medical Artificial Intelligence, Binzhou Medical University, Yantai 264003, China}
\affiliation{Beijing Advanced Innovation Center for Future Blockchain and Privacy Computing, Beihang University, Beijing 100191, China}
% \date{\today}% It is always \today,

\begin{abstract}
Reinforcement learning provides a framework for studying how individuals adjust their behavior through repeated interaction and feedback in social dilemmas. In Q-learning, exploration controls how often agents choose actions other than those favored by their current learned Q-values. Yet existing models usually treat the exploration rate as a constant parameter. In systems with social evaluation, however, trial-and-error behavior carries different costs and opportunities for agents with different reputations, making exploration dependent on social standing rather than uniform across agents. Herein, we develop a spatial Prisoner’s Dilemma model in which Q-learning agents adapt their exploration rates according to local reputation differences, while reputation is updated through an asymmetric, state-dependent rule. Results show that adaptive exploration and asymmetric reputation updating each promote cooperation, but their combination produces a stronger reinforcing effect than either mechanism alone. Low-reputation agents explore more and can recover reputation through cooperation, while high-reputation agents explore less and avoid reputation losses caused by defection. This mechanism also reorganizes cooperation in space, producing a stable checkerboard-like coexistence at intermediate reputation concern. In addition, cooperation is most vulnerable at intermediate baseline exploration rates, whereas stronger asymmetric reputation updating mitigates this exploration-induced disruption. These results suggest that reputation can act not only as a record of past behavior, but also as a dynamic signal that regulates exploratory behavior during learning and thereby stabilizes cooperation.
\end{abstract}

\maketitle

\begin{quotation}
Reputation is a central mechanism for explaining cooperation, but in most evolutionary models it enters the dynamics after behavior has been observed, serving as a social record for evaluation, interaction, or subsequent strategy choice. This study advances this view by showing that reputation can also act before action selection by regulating exploration in reinforcement learning. We develop a spatial Prisoner’s Dilemma model that couples reputation-based exploration with asymmetric, state-dependent reputation updating. The results show that this coupling promotes cooperation and reshapes its spatial organization. By endogenizing exploration through social information, this work provides a framework for understanding how cooperation emerges when social evaluation shapes not only how actions are judged, but also how behavioral alternatives are generated during learning.
\end{quotation}

\section{\label{sec:level1}INTRODUCTION}

Cooperation is widespread in biological systems and human societies~\cite{rand2013human,axelrod1981evolution}, yet it is difficult to explain from the perspective of Darwinian selection because individually beneficial actions can undermine collective welfare~\cite{sigmund2010calculus}. This conflict is formalized as a social dilemma~\cite{van2014social}, and motivates the question of how cooperation can emerge and persist among self-interested competitors~\cite{pennisi2005did}. Evolutionary game theory (EGT)~\cite{smith1973logic,taylor1978evolutionary} provides a theoretical framework for addressing this question by linking interaction structures~\cite{ohtsuki2006simple,perc2010coevolutionary,perc2017statistical,wang2026public,zhu2024evolutionary}, payoff incentives~\cite{wang2024evolutionary,wang2026survival}, and behavioral update rules~\cite{wang2023evolution,wang2023inertia,wang2023conflict,wang2023greediness}. Canonical models such as the Prisoner’s Dilemma game (PDG) capture the tension between short-term individual advantage and long-term collective welfare~\cite{axelrod1980effective,szabo1998evolutionary}.

Over decades of research, many mechanisms have been shown to promote cooperation. These include kin selection, direct reciprocity, indirect reciprocity, group selection, and spatial reciprocity~\cite{nowak2006evolutionary}. Cooperation can also be reinforced by institutional incentives such as reward and punishment~\cite{sigmund2001reward,szolnoki2010reward,szolnoki2011phase,zhu2023effects,han2024evolutionary,wang2024evolutionary} and by factors like aspiration~\cite{zhou2021aspiration,chen2024cooperation} or environmental feedback~\cite{weitz2016oscillating,tilman2020evolutionary,wang2020eco}. In social settings, cooperation also depends on how individuals are evaluated and remembered. Reputation allows individuals to condition their behavior on others’ past actions, thereby influencing future opportunities for cooperation~\cite{fu2008reputation,santos2018social,xia2023reputation,wang2023reputation}. In models of indirect reciprocity, reputation is updated by assessment rules that map observed actions to a public score~\cite{ohtsuki2004should,ohtsuki2006leading,hilbe2018indirect,wei2025indirect}. A common baseline is first-order assessment, where cooperation increases reputation and defection decreases it~\cite{nowak1998evolution,nowak2005evolution}.

How agents adapt their behavior is crucial in dynamic environments, because individuals do not know the optimal strategy in advance. Instead, they learn from repeated interactions and adjust their decisions based on feedback. This challenge motivates integrating EGT with multi-agent reinforcement learning (MARL)~\cite{watkins1992q,sutton2018reinforcement,koster2025deep,mckee2023scaffolding,wang2022levy,fan2022incorporating,geng2022reinforcement,xu2024reinforcement,mintz2025evolutionary,xie2026reinforcement,hou2017evolutionary,li2025memory}. Recent reviews have summarized the rapid development of evolutionary game dynamics under the reinforcement-learning paradigm~\cite{zheng2026brief}. Applications now span the evolution of cooperation in social dilemmas~\cite{ zhao2025evolution}, resource coordination and allocation~\cite{ zheng2025optimal,zhang2026dual}, and species coexistence in ecological systems~\cite{jiang2026decoding}. Related work has also shown that the way agents explore alternative actions can qualitatively reshape evolutionary outcomes~\cite{zhu2026evolutionary}. Together, these studies demonstrate that collective behavior depends not only on the underlying payoff structure, but also on how agents represent experience, evaluate future consequences, and explore behavioral alternatives.

A growing line of research has further incorporated reputation into learning-based evolutionary models and shown that social evaluation can promote cooperation~\cite{zou2024incorporating,ren2023reputation,xie2025reputation,ren2025bottom,zhu2025q,zhang2025q}. In these models, exploration is implemented as an algorithmic source of random action. Under a fixed exploration rate, an agent occasionally chooses an action that is not favored by its current learned values, and this deviation is treated in the same way for all individuals. Such a treatment is reasonable when exploration is viewed only as learning noise. In social dilemmas with reputation, however, exploratory behavior can also carry social meaning. The same behavioral deviation may have different costs and opportunities depending on an individual’s reputation and local standing. A low-reputation individual may use behavioral trial to rebuild trust, whereas a high-reputation individual may need to avoid unnecessary deviations that could damage accumulated trust. This suggests that reputation can regulate exploration during learning, rather than only affect payoff, fitness, or strategy evaluation after actions are taken~\cite{sutton2018reinforcement,tokic2011value,shen2024adaptive}.

Furthermore, to fully capture the coupling between exploration and reputation, it is important to consider how reputation itself is updated. Many score-based reputation models use a symmetric updating rule, where cooperation and defection change reputation by equal amounts in opposite directions~\cite{nowak1998evolution,nowak2005evolution,zhu2024reputation}. This simplifying assumption does not capture state-dependent tolerance and forgiveness, since a given action has the same reputational effect regardless of the actor’s prior reputation. However, evidence from social psychology shows that evaluations can be asymmetric and depend on observers’ expectations and prior impressions~\cite{skowronski1989negativity,fiske2018social,baumeister2001bad,lim2023trust}. For example, a high-status individual may be held to a stricter standard, so even a single norm violation can cause a disproportionately large loss of reputation. In contrast, a low-status individual might face persistent distrust, or they might be more readily forgiven if observers reward reparative behavior~\cite{fragale2009higher,dong2019second,chen2025impact}. 

Motivated by these considerations, we propose a spatial PDG model that couples Q-learning with two reputation-based processes. First, agents adapt their exploration rates according to the difference between their own reputation and the average reputation of their neighbors. Second, reputation is updated by an asymmetric, state-dependent rule, so the reputational effect of cooperation or defection depends on the agent’s prior standing. In this framework, reputation influences behavior before an action is chosen by regulating exploration, while the chosen action reshapes reputation after social evaluation. This feedback allows us to examine how reputation-based adaptive exploration and asymmetric reputation updating jointly shape the evolution of cooperation in structured populations.

Our simulations show that reputation-based adaptive exploration and asymmetric reputation updating reinforce each other in promoting cooperation. Compared with the fixed-exploration baseline, cooperation is enhanced when exploration is directed more toward low-reputation agents and less toward high-reputation agents. This pattern allows low-reputation agents to recover standing through cooperative trials, while reducing reputation losses caused by defective deviations among high-reputation agents. The effect is further strengthened by asymmetric reputation updating, which assigns larger gains to low-reputation cooperators and larger losses to high-reputation defectors. Beyond increasing cooperation, the joint mechanism changes the spatial organization of the population. At intermediate reputation concern, the system forms a stable checkerboard-like coexistence of high-reputation cooperators and low-reputation defectors, whereas stronger reputation concern drives the system toward near-full cooperation. We also find that cooperation is most vulnerable at intermediate baseline exploration rates, and that stronger asymmetric reputation updating mitigates this exploration-induced disruption.

Taken together, these findings broaden the role of reputation in learning-based evolutionary games from an evaluative record to a factor that conditions learning dynamics. When exploration depends on social standing, trial-and-error behavior is no longer a uniform perturbation imposed on all agents, but becomes socially differentiated. Reputation therefore affects cooperation not only by changing the consequences of behavior, but also by shaping the behavioral possibilities that learning can explore. Our work therefore provides a framework for studying how social information becomes embedded in adaptive dynamics and thereby reshapes the emergence and stability of cooperation.

\section{MODEL}\label{sec2}
\subsection{Spatial Prisoner's Dilemma Game}
We consider a population of agents on an $L \times L$ square lattice with periodic boundary conditions. Each lattice site hosts a single agent. Interaction topology is defined by a von Neumann neighborhood, meaning each agent interacts with its four nearest neighbors. At each interaction step, every agent plays the PDG with each of its neighbors and each pairwise interaction yields a payoff according to the payoff matrix and strategy choices.

Each agent has two possible strategies: Cooperation (C) or Defection (D). The payoff for an interaction is determined by a matrix $\mathbf{M}$, with entries $(R, S; T, P)$ following the canonical ordering $T > R > P > S$ and $2R > T+S$. Mutual cooperation yields $R$ for both players, mutual defection yields $P$, and a defector against a cooperator receives $T$ while the cooperator receives $S$. In this study, we adopt the weak PDG parametrization~\cite{nowak1992evolutionary}, setting $R=1$, $P=S=0$, and $T=b$, where $1 < b < 2$. The payoff matrix $\mathbf{M}$ is thus:
\begin{equation}
    \mathbf{M} = \begin{pmatrix} R & S \\ T & P \end{pmatrix} = \begin{pmatrix} 1 & 0 \\ b & 0 \end{pmatrix}.
\end{equation}

The strategy of an agent $i$ at time $t$ is represented by a basis vector, where $\mathbf{s}_i = (1, 0)^{\mathsf{T}}$ corresponds to cooperation and $\mathbf{s}_i = (0, 1)^{\mathsf{T}}$ corresponds to defection. The total payoff accrued by agent $i$ at time $t$, denoted $P_i(t)$, is the sum of payoffs from games with each of its neighbors:
\begin{equation}
    P_i(t) = \sum_{j \in \Omega_i} \mathbf{s}_i(t)^{\mathsf{T}} \mathbf{M} \mathbf{s}_j(t),
\label{eq:payoff}
\end{equation}
where $\Omega_i$ denotes the set of neighbors for agent $i$.

\subsection{Asymmetric Reputation Dynamics}
To model social evaluation, we assign every agent a reputation score that updates over time in an asymmetric manner. Let $R_i(t)$ be the reputation of agent $i$ at time $t$. The update of $R_i$ depends on agent $i$’s action $s_i(t)$ (C or D) and its previous reputation $R_i(t-1)$. We define a reputation threshold $A$ that divides agents into low-reputation ($R_i < A$) and high-reputation ($R_i \ge A$) categories. The reputation update rule is formulated as follows:
\begin{equation}
    R_i(t) = 
    \begin{cases} 
        R_i(t-1) + \delta, & \text{if } s_i(t) = \text{C and } R_i(t-1) < A, \\
        R_i(t-1) + 1, & \text{if } s_i(t) = \text{C and } R_i(t-1) \ge A, \\
        R_i(t-1) - \delta, & \text{if } s_i(t) = \text{D and } R_i(t-1) \ge A, \\
        R_i(t-1) - 1, & \text{if } s_i(t) = \text{D and } R_i(t-1) < A,
    \end{cases}
\label{eq:reputation}
\end{equation}
Here, $\delta>0$ is the reputation sensitivity parameter governing the state dependence of reputation updating. When $\delta=1$, Eq.~(\ref{eq:reputation}) reduces to a symmetric benchmark: cooperation increases reputation by one and defection decreases reputation by one, regardless of whether the agent's prior reputation is below or above $A$. For $\delta>1$, reputation updating becomes more punitive toward high-reputation defectors and more rewarding toward low-reputation cooperators. This asymmetric design is motivated by evidence from social psychology that evaluations depend on observers' prior expectations: high-reputation individuals are generally held to stricter standards, so a defection can produce a disproportionately large reputation loss~\cite{fragale2009higher,baumeister2001bad}, whereas cooperation by a low-reputation individual may be interpreted as reparative behavior and receive a stronger reputational reward~\cite{skowronski1989negativity, fiske2018social}. When $0<\delta<1$, the direction of the asymmetry is reversed, such that low-reputation cooperators receive smaller reputation gains and high-reputation defectors incur smaller reputation losses than under the symmetric benchmark. This regime captures the possibility that an established good record buffers high-reputation individuals against reputational damage~\cite{fiske2018social}. The parameter $\delta$ therefore allows the model to represent different regimes of expectation-dependent social evaluation.

Reputation is assumed to be nonnegative and bounded, reflecting a finite evaluation scale. We therefore restrict $R_i(t)$ to $[R_{\min},R_{\max}]$ (with $R_{\min}\ge 0$) and choose the threshold consistently within the same range, $A\in(R_{\min},R_{\max})$; in simulations, we enforce the bounds by clipping $R_i(t)$ after each update.

\subsection{Fitness Calculation}
In real-world societies, game payoffs represent the immediate material returns obtained from current interactions, whereas social standing, reflected by reputation, captures longer-term advantages by influencing access to cooperative partners, reciprocal relationships, and future resource opportunities. High-reputation individuals are therefore more likely to secure sustained benefits over time~\cite{fu2008reputation,milinski2002reputation,santos2018social}. Accordingly, individual fitness should reflect both material payoff and social standing. We therefore define each agent's fitness as a weighted combination of its game payoff and reputation~\cite{zhu2022exposure}. Specifically, the fitness of agent $i$ at time $t$ is given by a weighted sum of its total payoff and normalized reputation:
\begin{equation}
f_i(t) = (1-\theta) P_i(t) + \theta \frac{R_i(t)-R_{\min}}{R_{\max}-R_{\min}}4b,
\label{eq:fitness}
\end{equation}
where $\theta \in [0,1]$ is a weight capturing the agent’s concern for reputation. When $\theta = 0$, fitness depends only on payoff, whereas $\theta = 1$ means only reputation matters; intermediate values blend the two. Because the normalized reputation $(R_i(t)-R_{\min})/(R_{\max}-R_{\min})$ lies in $[0,1]$, multiplying it by $4b$ maps the unweighted reputational component to the interval $[0,4b]$. In the weak PDG on a square lattice with four nearest neighbors, the one-round game payoff satisfies $0\leq P_i(t)\leq4b$, with the maximum attained when a defector interacts with four cooperative neighbors. The two components are therefore placed on the same numerical range before the weights $1-\theta$ and $\theta$ are applied.

\begin{algorithm}[H]
    \caption{Q-learning with reputation-based adaptive exploration}
    \label{alg:algorithm}
    \begin{algorithmic}[1]
    \STATE \textbf{Input:} Lattice size $L$ ($N=L^2$), total Monte Carlo steps $T_{\mathrm{MCS}}$, parameters $b,\delta,\eta,\theta,\varepsilon_0,A,R_{\min},R_{\max},\alpha,\gamma$.
    \STATE \textbf{Output:} Stationary fraction of cooperators $\rho_{\mathrm{C}}$
    \STATE \textbf{Initialize:} For all agents $i\in\{1,\dots,N\}$, set $Q_i(s,a)=0$ for $s\in\mathcal{S},a\in\mathcal{A}$; set $R_i(0)=A$; assign initial state $s_i(0)\in\{\text{C},\text{D}\}$ uniformly at random.
    \FOR{$t=1$ to $T_{\mathrm{MCS}}$} 
        \FOR{$k=1$ to $N$}
            \STATE Randomly select one agent $i\in\{1,\dots,N\}$.
            \STATE Compute $\varepsilon_i(t)$ by Eq.~(\ref{eq:exploration})
            \STATE Let $s \leftarrow s_i$ (agent $i$'s current state, i.e., its previous action).
            \IF{a uniform random number $p < \varepsilon_i(t)$}
                \STATE Select action $a$ uniformly at random from $\mathcal{A}=\{\text{C},\text{D}\}$.
            \ELSE
                \STATE Select action $a \in \arg\max_{a'\in\mathcal{A}} Q_i(s,a')$.
            \ENDIF
            \STATE Set $\mathbf{s}_i(t)$ according to $a$ and compute $P_i(t)$ by Eq.~(\ref{eq:payoff}).
            \STATE Update reputation $R_i$ by Eq.~(\ref{eq:reputation}) and clip it to $[R_{\min},R_{\max}]$.
            \STATE Compute fitness $f_i$ by Eq.~(\ref{eq:fitness}).
            \STATE Update $Q_i(s,a)$ by Eq.~(\ref{eq:qupdate}) with $s' \leftarrow a$.
            \STATE Update state: $s_i \leftarrow a$.
        \ENDFOR
    \ENDFOR
    \end{algorithmic}
\end{algorithm}

\subsection{Q-Learning Framework}
Each agent is modeled as an independent reinforcement learning player that seeks to maximize its long-term fitness. We implement this via a self-interested Q-learning algorithm~\cite{wang2022levy,fan2022incorporating}, where each agent learns from its own experience. The strategic decision process for each agent $i$ can be viewed as a Markov Decision Process (MDP) with state space $\mathcal{S}$ and action space $\mathcal{A}$. The state is defined by the agent’s previous action, so $\mathcal{S}=\{\text{C},\text{D}\}$, and the action space is $\mathcal{A}=\{\text{C},\text{D}\}$.

Agent $i$ maintains an action-value function $Q_i(s,a)$ for each state-action pair, which estimates the expected cumulative future fitness if the agent is currently in state $s$ and then takes action $a$. These values are stored in a $2 \times 2$ Q-table for each agent:
\begin{equation}
    \mathbf{Q}_i = \begin{pmatrix} Q_i(\text{C}, \text{C}) & Q_i(\text{C}, \text{D}) \\ Q_i(\text{D}, \text{C}) & Q_i(\text{D}, \text{D}) \end{pmatrix},
\end{equation}
where, for example, $Q_i(\text{D}, \text{C})$ is the Q-value if agent $i$’s last action was D and it chooses C now. Agents update these Q-values based on the outcomes of interactions. We employ an $\epsilon$-greedy policy for action selection: with probability $1 - \epsilon_i(t)$, agent $i$ chooses the action with the highest $Q_i(s,a)$ for its current state $s$ (exploitation), and with probability $\epsilon_i(t)$, it selects a random action (exploration). After agent $i$ takes action $a$ in state $s$ and obtains a fitness reward $f_i(t)$, it updates its Q-value for $(s,a)$ using the standard Q-learning rule:
\begin{equation}
    Q_i(s, a) \leftarrow Q_i(s, a) + \alpha [f_i(t) + \gamma \max_{a'} Q_i(s', a') - Q_i(s, a)],
\label{eq:qupdate}
\end{equation}
where $s$ was the state before taking $a$, and $s’$ is the new state after the action (in our formulation, $s’ = a$, since the agent’s next state is its current action). Here $\alpha \in (0,1]$ is the learning rate and $\gamma \in [0,1)$ is the discount factor accounting for future rewards.

\subsection{Reputation-Based Adaptive Exploration Rate}
Unlike models with a fixed exploration probability, we let an agent’s exploration rate $\epsilon_i(t)$ adapt dynamically based on its social context. We modulate $\epsilon_i(t)$ according to the difference between agent $i$’s reputation and the average reputation of its neighbors. Let $\bar{R}_{\Omega_i}(t)$ denote the mean reputation of the neighbors of $i$. We define the adaptive exploration rate as:
\begin{equation}
\varepsilon_i(t) = \varepsilon_0^{1 + \tanh\left[\eta\left(\frac{R_i(t) - \bar{R}_{\Omega_i}(t)}{R_{\max} - R_{\min}}\right)\right]},
\label{eq:exploration}
\end{equation}
where $\varepsilon_0\in[0,1]$ is the baseline exploration rate and $\eta\in[-1,1]$ controls how relative reputation biases exploration. When $\eta>0$, agents with lower reputation than their neighborhood average explore more, while higher-reputation agents explore less; $\eta<0$ reverses this tendency. Setting $\eta=0$ yields $\varepsilon_i(t)=\varepsilon_0$, recovering the fixed exploration case.

\subsection{Parameter Configuration}
We employ an asynchronous update scheme in our simulations. One full Monte Carlo step (MCS) consists of $L^2$ elementary steps, and each elementary step randomly selects one agent to update according to Algorithm~\ref{alg:algorithm}. We run simulations for $1\times10^5$ MCS and collect statistics by averaging over the last $5{,}000$ MCS. Each data point is further averaged over 20 independent runs. Table~\ref{tab:params} summarizes the model parameters.

\begin{table}[h!]
\centering
\caption{Model parameters and their descriptions}
\label{tab:params}
\begin{tabular}{@{}ll@{}}
\toprule
\textbf{Symbol} & \textbf{Description} \\
\midrule
$L$ & Lattice dimension ($L \times L$ grid) \\
$b$ & Temptation to defect in the PDG \\
$R_{\min}$ & Minimum reputation value \\
$R_{\max}$ & Maximum reputation value \\
$A$ & Reputation threshold for high/low status \\
$\delta$ & Reputation sensitivity parameter \\
$\theta$ & Reputation concern (weight in fitness) \\
$\alpha$ & Learning rate for Q-table updates \\
$\gamma$ & Discount factor for future rewards \\
$\epsilon_0$ & Baseline exploration rate \\
$\eta$ & Exploration bias based on reputation difference \\
\bottomrule
\end{tabular}
\end{table}

\begin{figure*}[htbp]
\centering
\includegraphics[width=0.9\textwidth]{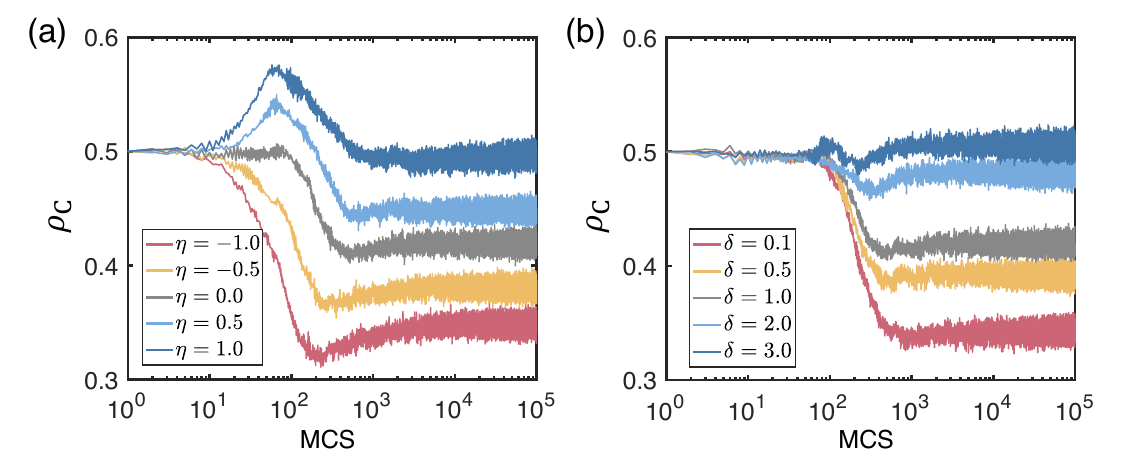}
\caption{Adaptive exploration and asymmetric reputation updating independently and directionally reshape the evolution of cooperation. (a) Time evolution of the $\rho_{\mathrm{C}}$ for different $\eta$ under symmetric reputation updating ($\delta=1$). When $\eta > 0$, cooperation is promoted; conversely, when $\eta < 0$, cooperation is inhibited. (b) Time evolution of the $\rho_{\mathrm{C}}$ for different $\delta$ under fixed exploration ($\eta=0$, i.e., $\varepsilon_i(t)=\varepsilon_0$). When $\delta > 1$, cooperation is promoted, whereas $\delta < 1$ leads to a decline in cooperation levels. The fixed parameters are $b=1.6$, $\theta=0.6$, $\epsilon_0=0.02$.
}
\label{fig:fig1}
\end{figure*}

\begin{table}[htbp]
    \centering
    \caption{Notation for exploration and reputation mechanisms}
    \label{tab:strategies}
    \renewcommand{\arraystretch}{1.3} 
    \begin{tabular}{l l p{5.5cm}}
        \toprule
        \textbf{Symbol} & \textbf{Parameter} & \textbf{Description} \\
        \midrule
        \multicolumn{3}{l}{\textit{Exploration mechanism}} \\
        $\mathrm{E}^{0}$ & $\eta = 0$ & Fixed exploration rate (Baseline case). \\
        $\mathrm{E}^{-}$ & $\eta < 0$ & Low reputation $\to$ Low exploration.
        
        High reputation $\to$ High exploration. \\
        $\mathrm{E}^{+}$ & $\eta > 0$ & Low reputation $\to$ High exploration.
        
        High reputation $\to$ Low exploration. \\
        \midrule
        \multicolumn{3}{l}{\textit{Reputation update rule}} \\
        $\mathrm{R}^{0}$ & $\delta = 1$ & Symmetric reputation updates. \\
        $\mathrm{R}^{-}$ & $\delta < 1$ & Weaker reputation gain for low-reputation cooperation and weaker reputation loss for high-reputation defection. \\
        $\mathrm{R}^{+}$ & $\delta > 1$ & Stronger reputation gain for low-reputation cooperation and stronger reputation loss for high-reputation defection.  \\
        \bottomrule
    \end{tabular}
\end{table}

\section{Analysis of Results}\label{sec3}

In the simulations, we fix $L=200$ throughout and confirm that enlarging the lattice does not change the stationary outcomes reported below. We also tested different initial strategy fractions and different initial reputation distributions, and found that these variations do not affect the stationary cooperation level or the qualitative phase behavior. Unless stated otherwise, we fix $R_{\min}=0$, $R_{\max}=100$, and $A=50$. To verify that the reported results are not tied to these particular numerical choices, we further examine four different reputation ranges and four different reputation thresholds in Appendix~\ref{app:reputation_robustness}. Although varying these parameters produces modest quantitative changes in the stationary cooperation level, the directional effect of reputation-based adaptive exploration and its reinforcement by asymmetric reputation updating remain unchanged. In addition, for comparability with prior learning-based evolutionary studies~\cite{zou2024incorporating,ren2023reputation,zhu2025q,xie2025reputation,fan2022incorporating}, we set $\alpha=0.8$ and $\gamma=0.8$ in the main simulations. To assess whether the conclusions depend on these particular learning parameters, Appendix~\ref{app:learning_parameters} examines the baseline, the two individual mechanisms, and their combination over the $(\alpha,\gamma)$ parameter plane.

\begin{figure*}[htbp]
\centering
\includegraphics[width=1\textwidth]{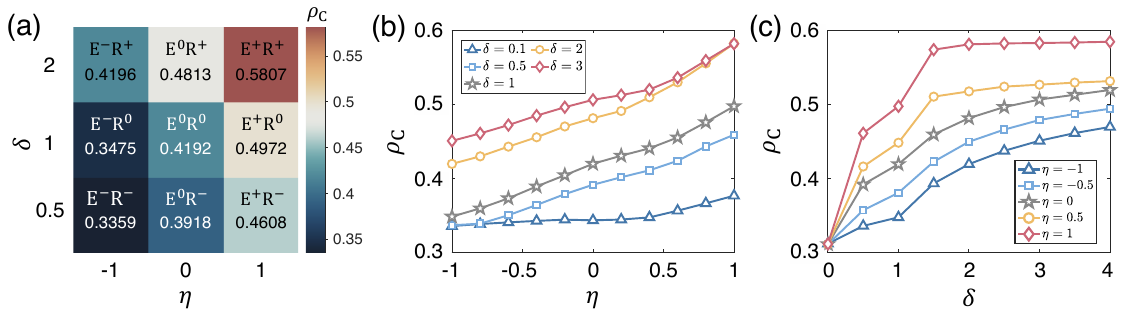}
\caption{Synergistic effect between adaptive exploration and asymmetric reputation. (a) Heat map of the fraction of cooperation $\rho_{\mathrm{C}}$ for the nine combinations of exploration mechanism (columns, controlled by $\eta$) and reputation update rule (rows, controlled by $\delta$), where each cell reports the corresponding $\rho_{\mathrm{C}}$. (b) Fraction of cooperation $\rho_{\mathrm{C}}$ as a function of $\eta$ for different $\delta$ values, showing that the cooperative advantage of $\eta>0$ becomes stronger as $\delta$ increases (i.e., asymmetric reputation updating amplifies the effect of reputation-directed exploration). (c) Fraction of cooperation $\rho_{\mathrm{C}}$ as a function of $\delta$ for different $\eta$ values, showing that increasing $\delta$ promotes cooperation, but as $\delta$ continues to increase, the marginal gain in the frequency of cooperation decreases. The fixed parameters are $b=1.6$, $\theta=0.6$, $\epsilon_0=0.02$.
}
\label{fig:fig2}
\end{figure*}

\begin{figure*}[htbp]
\centering
\includegraphics[width=1\textwidth]{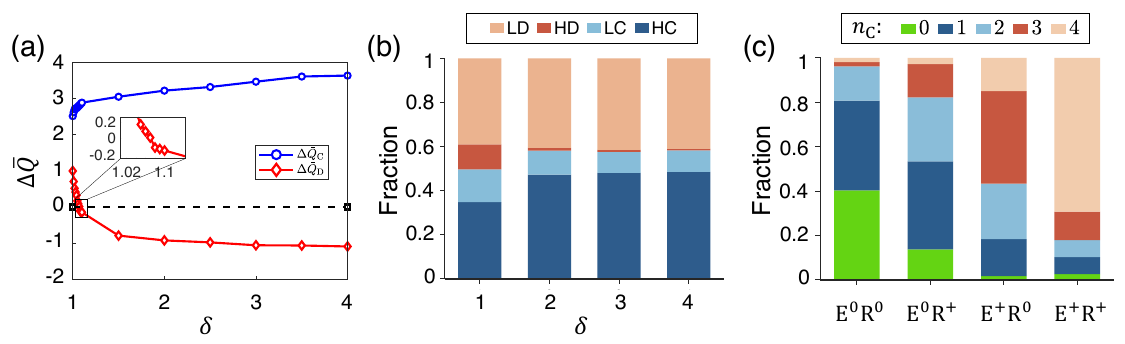}
\caption{
Microscopic evidence for how the joint mechanism stabilizes cooperation. (a) Steady-state Q-value gaps $\Delta\bar{Q}$ as a function of $\delta$ under the exploration bias ($\eta=1$). Here, $\Delta\bar{Q}_{\mathrm{C}}>0$ indicates that agents who previously cooperated value continuing to cooperate more than switching to defection, while $\Delta\bar{Q}_{\mathrm{D}}>0$ indicates that agents who previously defected value switching to cooperation more than persisting in defection. (b) Fractions of the four behavioral--reputational types (LC/HC for low-/high-reputation cooperators; LD/HD for low-/high-reputation defectors) versus $\delta$ under $\eta=1$. (c) Distribution of the number of cooperative neighbors $n_\mathrm{C}\in\{0,1,2,3,4\}$ for cooperation survival events (a $\mathrm{D}\!\to\!\mathrm{C}$ switch followed by at least two further consecutive cooperative actions) under four representative mechanisms. The fixed parameters are $b=1.6$, $\theta=0.6$, $\epsilon_0=0.02$.
}
\label{fig:fig3}
\end{figure*}

\begin{figure*}[htbp]
\centering
\includegraphics[width=1\textwidth]{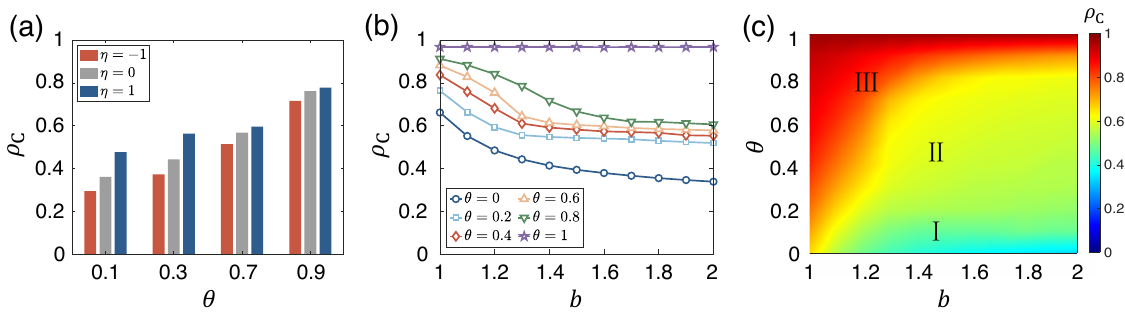}
\caption{Reputation concern governs the cooperation regime. (a)  Bar chart of the fraction of cooperation $\rho_{\mathrm{C}}$ versus $\theta$ under three exploration biases $\eta\in\{-1,0,1\}$ at $b=1.6$. $\rho_{\mathrm{C}}$ increases with $\theta$, while the differences among $\eta$ shrink at large $\theta$. (b) Fraction of cooperation $\rho_{\mathrm{C}}$ as a function of $b$ for $\theta\in\{0,0.2,0.4,0.6,0.8,1\}$ at $\eta=1$. Reputation entering fitness ($\theta>0$) markedly elevates $\rho_{\mathrm{C}}$. Intermediate $\theta$ yields a saturation state near $\rho_{\mathrm{C}}\approx0.6$. (c) Heat map of $\rho_{\mathrm{C}}$ in the $(b,\theta)$ plane at $\eta=1$. The lower region (I) corresponds to low cooperation, the middle region (II) shows a saturation state with $\rho_{\mathrm{C}}\approx0.6$, and the upper region (III) exhibits high cooperation with $\rho_{\mathrm{C}}>0.6$. Dashed curves indicate the boundaries between these regimes. The fixed parameters are $\delta=3$, $\epsilon_0=0.02$.
}
\label{fig:fig4}
\end{figure*}

\begin{figure*}[htbp]
\centering
\includegraphics[width=0.9\textwidth]{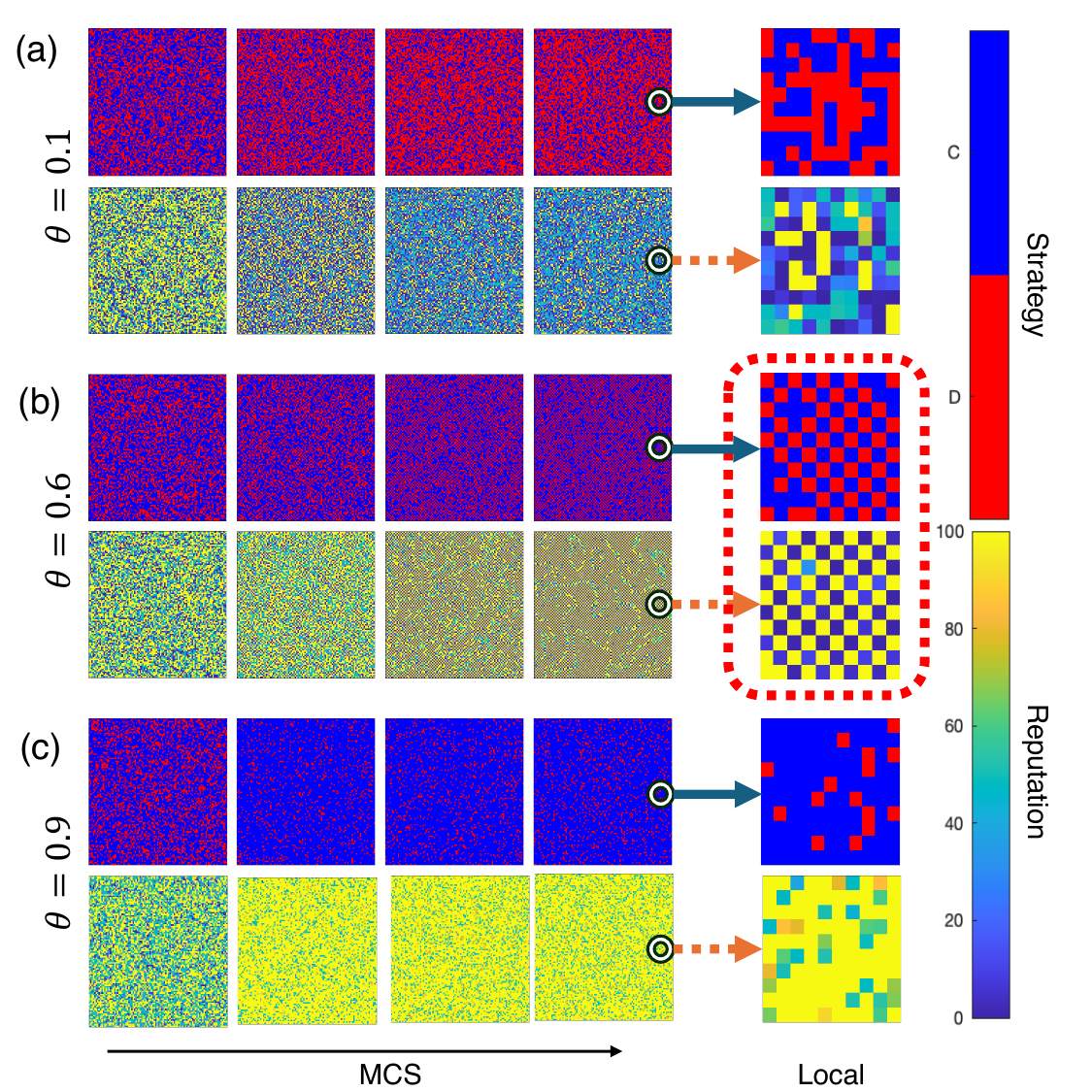}
\caption{Spatiotemporal evolution of strategy and reputation for different reputation concern. Snapshots of strategy (top row in each block; C in blue and D in red) and reputation (bottom row in each block; color scale) at $\theta\in\{0.1,0.6,0.9\}$. Columns show increasing MCS from left to right, and the right panels show a local view taken from the marked area in the final snapshot. (a) $\theta=0.1$ yields dominant defection with small cooperative clusters and generally low reputation. (b) $\theta=0.6$ yields stable coexistence in which high-reputation cooperators and low-reputation defectors occupy interwoven neighboring sites, producing a checkerboard-like pattern in strategy and a corresponding local ordering in reputation. (c) $\theta=0.9$ yields near full cooperation with sparse defectors and high reputation for most agents. The fixed parameters are $\delta=3$, $\eta=1$, $b=1.5$ and $\epsilon_0=0.02$.
}
\label{fig:fig5}
\end{figure*}

\begin{figure}[htbp]
\centering
\includegraphics[width=0.5\textwidth]{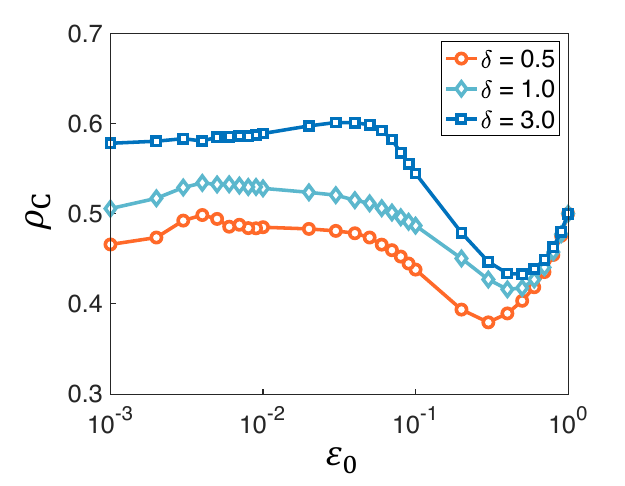}
\caption{Impact of baseline exploration rate. We show the fraction of cooperation $\rho_{\mathrm{C}}$ as a function of the baseline exploration rate $\epsilon_0$ for $\delta\in\{0.5,1,3\}$. The fraction of cooperation increases at small $\epsilon_0$, decreases over an intermediate range, then rises toward $\rho_{\mathrm{C}}\approx0.5$ as $\epsilon_0\to1$. Asymmetric updating with $\delta>1$ reduces the cooperation drop at intermediate $\varepsilon_0$, whereas $\delta<1$ enlarges it. The fixed parameters are $\eta=1$, $\theta=0.6$, $b=1.6$.
}
\label{fig:fig6}
\end{figure}

\subsection{Separate Effects of Adaptive Exploration and Asymmetric Reputation Updating}

To isolate the roles of adaptive exploration and asymmetric reputation updating, we vary one mechanism at a time. Specifically, we fix $\delta=1$ in Fig.~\ref{fig:fig1}(a) to remove asymmetry in reputation updating, and we fix $\eta=0$ in Fig.~\ref{fig:fig1}(b) to remove reputation dependence in exploration.

Figure~\ref{fig:fig1}(a) shows the evolution of $\rho_{\mathrm{C}}$ for different exploration bias $\eta$ under symmetric reputation updating ($\delta=1$). When $\eta=0$, the model reduces to standard $\varepsilon$-greedy learning with a constant exploration rate $\varepsilon_0$. For $\eta>0$, agents with lower reputation than their neighborhood average explore more, while higher-reputation agents explore less. In this regime, the stationary cooperation level increases with $\eta$. In contrast, for $\eta<0$ the exploration pattern is reversed, and the stationary $\rho_{\mathrm{C}}$ decreases as $\eta$ becomes more negative. These results show that adaptive exploration affects cooperation, and the sign of $\eta$ determines whether the effect is cooperative or detrimental. To determine whether the effect of adaptive exploration is generated solely by the direct contribution of reputation to fitness, we additionally repeat the exploration comparison at $\theta=0$. As shown in Appendix~\ref{app:mechanism_isolation} (Fig.~\ref{fig:theta0}(a)), $\eta>0$ still increases the stationary cooperation level relative to the fixed-exploration case $\eta=0$, whereas $\eta<0$ reduces it. Therefore, reputation-based adaptive exploration retains its directional effect when reputation serves only as a signal for regulating exploration and does not contribute directly to fitness.

Figure~\ref{fig:fig1}(b) shows the evolution of $\rho_{\mathrm{C}}$ for different asymmetry levels $\delta$ under fixed exploration ($\eta=0$). The case $\delta=1$ corresponds to symmetric reputation updating. When $\delta>1$, cooperation produces a larger reputation increase for low-reputation agents, and defection produces a larger reputation decrease for high-reputation agents. Under this incentive structure, $\rho_{\mathrm{C}}$ converges to a higher stationary level, and the increase is stronger for larger $\delta$. When $0<\delta<1$, these reputation incentives are weakened, and the stationary cooperation level declines.

In summary, both mechanisms have a directional effect on cooperation. Cooperation is enhanced when exploration is concentrated on low-reputation agents ($\eta>0$) or when reputation updating strengthens rewards for low-reputation cooperation and penalties for high-reputation defection ($\delta>1$).

\subsection{\label{sec:level2} Synergistic Effect Between Adaptive Exploration and Asymmetric Reputation Updating}

We next examine the joint effects of reputation-based exploration and asymmetric reputation updating. For clarity, Table~\ref{tab:strategies} summarizes the notation used for the exploration mechanism ($\mathrm{E}$) and the reputation update rule ($\mathrm{R}$).

The combined outcomes across the nine settings are summarized in Fig.~\ref{fig:fig2}(a). Relative to the baseline $\mathrm{E}^{0}\mathrm{R}^{0}$, increasing $\eta$ alone ($\mathrm{E}^{+}\mathrm{R}^{0}$) or increasing $\delta$ alone ($\mathrm{E}^{0}\mathrm{R}^{+}$) raises $\rho_{\mathrm{C}}$. When both are applied together, cooperation increases further. We find that $\rho_{\mathrm{C}}$ under $\mathrm{E}^{+}\mathrm{R}^{+}$ exceeds both $\mathrm{E}^{+}\mathrm{R}^{0}$ and $\mathrm{E}^{0}\mathrm{R}^{+}$. This ranking shows that the two mechanisms reinforce each other rather than acting as substitutes. The same ranking persists when reputation is excluded from fitness. As shown in Appendix~\ref{app:mechanism_isolation} (Fig.~\ref{fig:theta0}(b)), when $\theta=0$, $\mathrm{E}^{+}\mathrm{R}^{+}$ still yields a higher cooperation level than $\mathrm{E}^{+}\mathrm{R}^{0}$ and $\mathrm{E}^{0}\mathrm{R}^{+}$.

To clarify where this reinforcement comes from, Fig.~\ref{fig:fig2}(b) and Fig.~\ref{fig:fig2}(c) examine the two control directions separately. For a fixed $\delta$, $\rho_{\mathrm{C}}$ increases with $\eta$, and the increase becomes stronger as $\delta$ grows (Fig.~\ref{fig:fig2}(b)), showing that asymmetric reputation updating amplifies the cooperative advantage of directing exploration toward low-reputation agents. Conversely, for a fixed $\eta$, $\rho_{\mathrm{C}}$ increases with $\delta$ (Fig.~\ref{fig:fig2}(c)). When $\eta>0$, $\rho_{\mathrm{C}}$ rises rapidly as $\delta$ crosses 1 and then levels off, so further increases in $\delta$ yield smaller gains. This motivates a microscopic analysis of how the joint mechanism reshapes learning incentives and population composition.

To explain the diminishing marginal gain in Fig.~\ref{fig:fig2}(c) for $\eta>0$ and $\delta>1$, we analyze the learning signals and the resulting population structure under the exploration bias ($\eta=1$).

Figure~\ref{fig:fig3}(a) tracks two Q-value gaps. Define
\begin{subequations}
    \begin{align}
        \Delta \bar{Q}_{\mathrm{C}}=\overline{Q(\mathrm{C},\mathrm{C})}-\overline{Q(\mathrm{C},\mathrm{D})},\\
        \Delta \bar{Q}_{\mathrm{D}}=\overline{Q(\mathrm{D},\mathrm{C})}-\overline{Q(\mathrm{D},\mathrm{D})},
    \end{align}
\end{subequations}
where the overline denotes an average over agents at steady state. A positive $\Delta \bar{Q}_{\mathrm{C}}$ indicates that cooperators assign higher value to persisting in cooperation than switching to defection, while a positive $\Delta \bar{Q}_{\mathrm{D}}$ indicates that defectors assign higher value to switching to cooperation than remaining in defection. As $\delta$ increases above 1, $\Delta \bar{Q}_{\mathrm{C}}$ grows and $\Delta \bar{Q}_{\mathrm{D}}$ decreases, so agents increasingly prefer to repeat their current action. Both curves then change more slowly as $\delta$ becomes large, which is consistent with the leveling-off behavior of $\rho_{\mathrm{C}}$.

The same situation is reflected in the population composition. As shown in Fig.~\ref{fig:fig3}(b), when $\delta=1$, HC, LC, HD, and LD all occupy non-negligible shares, indicating that reputation and strategy are not yet tightly coupled. When $\delta\ge 2$, the high-reputation group is dominated by cooperators and the low-reputation group is dominated by defectors, and the composition changes little with further increases in $\delta$. This pattern shows that the mechanism can reliably identify cooperators (defectors) and assign them high (low) reputation, consistent with social expectations. Once this correspondence is established, increasing $\delta$ mainly rescales the strength of the same separation, which explains why additional gains in $\rho_{\mathrm{C}}$ become limited.

Finally, Fig.~\ref{fig:fig3}(c) links the joint mechanism to cooperation stability under local temptation. Let $n_\mathrm{C}\in\{0,1,2,3,4\}$ be the number of cooperative neighbors of a focal agent. In the weak PDG, the immediate gain from defecting against cooperative neighbors increases with $n_C$ (a larger $n_\mathrm{C}$ corresponds to stronger temptation). We define a \emph{cooperation-survival event} as a transition $\mathrm{D}\to\mathrm{C}$ followed by at least two further consecutive cooperative actions. Fig.~\ref{fig:fig3}(c) shows the distribution of $n_C$ for these events under different mechanisms. Under $\mathrm{E}^{+}\mathrm{R}^{+}$, a large share of survival events occurs at $n_C=3$ or $4$, indicating that cooperation can persist even when the short-term incentive to defect is strong. In contrast, under $\mathrm{E}^{0}\mathrm{R}^{0}$ survival events concentrate at small $n_C$, meaning cooperation is mainly stable in low-temptation neighborhoods. This comparison supports the interpretation that $\mathrm{E}^{+}\mathrm{R}^{+}$ improves cooperation by stabilizing it under high temptation rather than by relying on sheltered local configurations.

\subsection{\label{sec:level3} Impact of the Reputation Concern}

We now examine how the reputation concern $\theta$, which weights reputation in fitness, shapes cooperation under the synergistic setting. As shown in Fig.~\ref{fig:fig4}(a), increasing $\theta$ raises the fraction of cooperation for all three exploration biases. Meanwhile, the differences among $\eta=-1,0,1$ shrink as $\theta$ increases. This indicates that when reputation contributes more to fitness, reputation-driven selection becomes the dominant force shaping behavior, and the additional effect introduced by the exploration bias becomes less pronounced.

The effect of $\theta$ becomes more evident when the temptation to defect increases. Figure~\ref{fig:fig4}(b) shows that introducing reputation into fitness ($\theta>0$) markedly improves cooperation compared with $\theta=0$. For $\theta=1$, cooperators occupy almost the whole population across the explored range of $b$. For intermediate values of $\theta$, cooperation decreases as $b$ increases and then stabilizes close to $\rho_{\mathrm{C}}\approx 0.6$, indicating a cooperation saturation state in which the long-run cooperation level becomes weakly sensitive to further increases in $b$.

These trends are summarized in the phase diagram in Fig.~\ref{fig:fig4}(c), where the $(b,\theta)$ plane can be divided into three representative regions. Region I corresponds to low cooperation, with $\rho_{\mathrm{C}}$ fluctuating around a relatively small value. Region II corresponds to the cooperation saturation state, where $\rho_{\mathrm{C}}$ stays around $0.6$ over a broad parameter range. Region III corresponds to high cooperation, with $\rho_{\mathrm{C}}$ exceeding $0.6$ and responding more strongly to changes in $b$ and $\theta$. Increasing $\theta$ expands Region III, whereas increasing $b$ compresses it and enlarges the saturation regime. This indicates that stronger reputation concern offsets the temptation to defect, while stronger temptation pushes the system toward coexistence rather than near-full cooperation.

To reveal the microscopic patterns behind the three regions, we fix $b=1.5$ and select $\theta=0.1$, $0.6$, and $0.9$, which correspond to Regions I--III in Fig.~\ref{fig:fig4}(c). Figure~\ref{fig:fig5} shows the spatiotemporal evolution of strategy and reputation.

For small $\theta$ (Fig.~\ref{fig:fig5}(a)), payoffs dominate fitness and reputation contributes little. Defectors expand by exploiting nearby cooperators, and the remaining cooperators survive mainly in small compact clusters. The reputation field drifts toward low values, consistent with the prevalence of defection. For intermediate $\theta$ (Fig.~\ref{fig:fig5}(b)), reputation and payoff jointly determine fitness, and the system evolves toward a stable spatial coexistence. High-reputation cooperators and low-reputation defectors occupy interwoven neighboring sites, producing a checkerboard-like organization in both strategy and reputation. The formation condition and the spatiotemporal organization of this structure are analyzed in Appendix~\ref{app:checkerboard}. The local fitness comparison explains why intermediate reputation concern can support anti-coordinated coexistence, while the state composition, nearest-neighbor statistics, checkerboard order parameters, spatial correlations, and transition correlations quantify its spatial ordering and temporal persistence. This spatial structure also supports the cooperation saturation level observed in Fig.~\ref{fig:fig4}(b) and Fig.~\ref{fig:fig4}(c). For large $\theta$ (Fig.~\ref{fig:fig5}(c)), reputation dominates fitness. Agents therefore learn to cooperate to maintain high reputation, and the population becomes nearly all cooperative. The remaining defectors are sparse and surrounded by cooperators, and their reputations stay low.

Overall, increasing $\theta$ strengthens the selective pressure induced by reputation, which raises cooperation and can drive the system from cluster-based survival, through a robust coexistence regime, to near-full cooperation.

\subsection{\label{sec:level4} Impact of the Baseline Exploration Rate}

Figure~\ref{fig:fig6} shows how the fraction of cooperation depends on the baseline exploration rate $\varepsilon_0$ under different asymmetry levels $\delta$. Across all cases, $\rho_{\mathrm{C}}$ changes non-monotonically as $\varepsilon_0$ increases. In the small-$\varepsilon_0$ range, cooperation rises slightly; at intermediate $\varepsilon_0$, cooperation drops markedly; and when $\varepsilon_0$ becomes close to 1, $\rho_{\mathrm{C}}$ increases again and approaches $0.5$.

When $\varepsilon_0$ is very small, action selection is nearly greedy and the dynamics are dominated by exploitation. A small increase in $\varepsilon_0$ introduces occasional trial moves, which helps agents correct early misjudgments and adjust their action values. As a result, $\rho_{\mathrm{C}}$ exhibits a mild upward trend, although the improvement remains limited in magnitude. As $\varepsilon_0$ enters an intermediate range, exploration becomes frequent enough to interfere with the formation of stable behavioral patterns. Random actions, in particular random defections, occur more often and disrupt local cooperative neighborhoods. This weakens the reputation--fitness feedback and leads to a pronounced decrease in $\rho_{\mathrm{C}}$. When $\varepsilon_0$ is very large, action choice is dominated by randomness and exploitation becomes ineffective. In this limit, neither cooperation nor defection can be consistently reinforced, and the population approaches an approximately unbiased mixed state with $\rho_{\mathrm{C}}\approx 0.5$.

The role of $\delta$ is reflected in both the overall level of cooperation and the position of the downturn. Larger $\delta$ maintains a higher $\rho_{\mathrm{C}}$ and shifts the onset of the decline to larger $\varepsilon_0$, indicating that stronger asymmetric reputation updating makes cooperative configurations more resistant to exploration-induced noise.

\section{Conclusion}\label{sec4}

Reinforcement learning provides a useful framework for modeling behavioral adaptation in social dilemmas, where agents update their actions through repeated interaction and feedback~\cite{watkins1992q,sutton2018reinforcement,koster2025deep}. In standard $\epsilon$-greedy learning, exploration represents occasional actions that are not selected by current learned values. When this probability is fixed, behavioral trial is generated at the same baseline rate for all agents and remains separate from reputation information. This provides a useful baseline, but it is restrictive in reputation-based interactions, where agents differ in their reputation relative to nearby agents. Allowing exploration to vary with relative reputation therefore makes action selection depend not only on learned values and past feedback, but also on the agent’s current social state before action selection.

In this work, we developed a spatial Prisoner’s Dilemma model that combines Q-learning with reputation-based exploration and asymmetric, state-dependent reputation updating. The exploration rate changes according to the difference between an agent’s reputation and the average reputation of its neighbors. Reputation also contributes to fitness together with payoff, and is updated according to both the current action and the agent’s prior reputation. Thus, reputation acts as a state variable with multiple roles: it affects fitness evaluation, regulates exploration, and is itself modified by behavior. This setting allows us to test whether making exploration reputation-dependent, together with state-dependent reputation updating, changes how cooperation emerges and is maintained in structured populations.

Our simulations show that both mechanisms promote cooperation, and that their combination yields a stronger effect than either mechanism alone. Cooperation is enhanced when exploration is biased toward low-reputation agents and reduced among high-reputation agents. This pattern supports reputation recovery through cooperative trials while limiting reputation-damaging exploratory defections among high-reputation agents. Asymmetric updating further strengthens cooperation when low-reputation cooperators receive larger reputation gains and high-reputation defectors receive larger reputation losses.

The analyses also show that the mechanism affects not only the level of cooperation, but also its spatial organization and robustness. Increasing the reputation weight can drive the system from low cooperation to a structured coexistence of high-reputation cooperators and low-reputation defectors, and eventually to near-full cooperation. Baseline exploration has a non-monotonic effect: cooperation is most vulnerable at intermediate exploration rates, and stronger asymmetric updating reduces this disruption. These results show that the cooperative effect of reputation-guided learning depends on how exploration is distributed and how reputation responds to cooperative and defective actions.

Taken together, this work shows that reputation can regulate the source and distribution of behavioral trial in learning populations. By making exploration reputation-dependent, the model turns trial-and-error from a fixed algorithmic perturbation into a socially structured component of adaptation. This provides a theoretical basis for explaining how cooperation can be strengthened when social information shapes not only the evaluation and fitness consequences of actions, but also the way behavioral alternatives are explored during learning. Future work can extend this approach by combining reputation-guided exploration with institutional incentives such as reward and punishment~\cite{sigmund2001reward,szolnoki2010reward,szolnoki2011phase}, or by replacing first-order scores with richer assessment rules from indirect reciprocity~\cite{ohtsuki2004should,ohtsuki2006leading,hilbe2018indirect}.

\section*{ACKNOWLEDGMENTS}
This work is supported by National Science and Technology Major Project (2022ZD0116800), Program of National Natural Science Foundation of China (12425114, 62141605, 12201026, 12301305, 62441617, 12501702), the Fundamental Research Funds for the Central Universities, Beijing Natural Science Foundation (Z230001), National Cyber Security-National Science and Technology Major Project (2025ZD1503700), the Opening Project of the State Key Laboratory of General Artificial Intelligence(Project No. SKLAGI2025OP16), and Beijing Advanced Innovation Center for Future Blockchain and Privacy Computing.

\section*{AUTHOR DECLARATIONS}
\subsection*{Conflict of Interest}
The authors have no competing interests to declare that are relevant to the content of this article.

\subsection*{Author Contributions}
\textbf{An Li:} Conceptualization (equal); Investigation (equal); Methodology (equal); Visualization (equal); Writing -- original draft (equal); Writing -- review \& editing (equal). \textbf{Wenqiang Zhu:} Conceptualization (equal); Investigation (equal); Methodology (equal); Visualization (equal); Writing -- original draft (equal); Writing -- review \& editing (equal). \textbf{Chaoqian Wang:} Methodology (equal); Writing -- review \& editing (equal). \textbf{Longzhao Liu:} Funding acquisition (equal); Supervision (equal). \textbf{Hongwei Zheng:} Investigation (equal). \textbf{Yishen Jiang:} Methodology (equal); Writing -- review \& editing (equal). \textbf{Xin Wang:} Conceptualization (equal); Funding acquisition (equal); Supervision (equal); Writing -- review \& editing (equal). \textbf{Shaoting Tang:} Funding acquisition (equal); Supervision (equal); Writing -- review \& editing (equal).

\section*{Data Availability Statement}
The data that support the findings of this study are available from the corresponding author upon reasonable request.

\appendix
\section{Robustness to Reputation Range and Threshold}
\label{app:reputation_robustness}

The main results are obtained with $R_{\min}=0$, $R_{\max}=100$, and $A=50$. Here, we examine whether the evolutionary outcomes depend on these specific choices.

\begin{figure*}[htbp]
\centering
\includegraphics[width=1\textwidth]{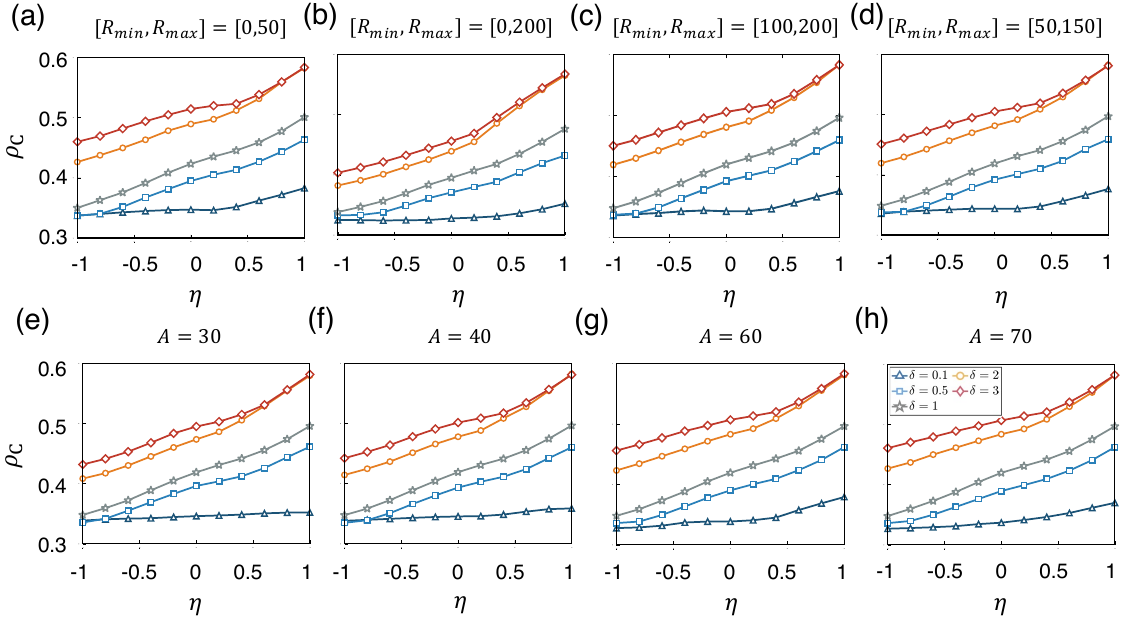}
\caption{
Robustness of cooperation to the reputation range and reputation threshold. (a)--(d) Fraction of cooperation $\rho_\mathrm{C}$ as a function of the exploration bias $\eta$ for $\delta\in\{0.1,0.5,1,2,3\}$ under four reputation ranges. For each range, the threshold $A$ and the initial reputation are set to the midpoint of the corresponding interval. (e)--(h) Fraction of cooperation $\rho_\mathrm{C}$ as a function of $\eta$ for the same values of $\delta$ under $A\in\{30,40,60,70\}$, with $R_{\min}=0$ and $R_{\max}=100$. In all cases, positive $\eta$ promotes cooperation, and this effect becomes stronger as $\delta$ increases. The fixed parameters are $b=1.6$, $\theta=0.6$, and $\varepsilon_0=0.02$.
}
\label{fig:reputation_robustness}
\end{figure*}

\subsection{Effect of the Reputation Range}
Figures~\ref{fig:reputation_robustness}(a)--(d) show the stationary fraction of cooperation $\rho_\mathrm{C}$ as a function of $\eta$ under four reputation ranges. For each range, the threshold $A$ and the initial reputation are set to the midpoint of the corresponding interval. Although the precise values of $\rho_\mathrm{C}$ vary slightly among the four panels, the qualitative relationships remain unchanged. In every case, cooperation increases as $\eta$ changes from negative to positive values, and this increase becomes stronger for larger $\delta$. Thus, adaptive exploration continues to promote cooperation, while asymmetric reputation updating reinforces this effect across all tested reputation ranges.

This robustness arises because both reputation-dependent components are defined relative to the reputation scale. The exploration rate depends on the normalized difference in Eq.~(\ref{eq:exploration}), while the reputation contribution to fitness is normalized by the same range width in Eq.~(\ref{eq:fitness}). Changing the reputation range therefore modifies how rapidly the fixed increments $1$ and $\delta$ move agents across the scale, producing moderate quantitative differences, but it does not reverse the directional effects of $\eta$ and $\delta$.

\subsection{Effect of the Reputation Threshold}
Figures~\ref{fig:reputation_robustness}(e)--(h) show the corresponding results for $A\in\{30,40,60,70\}$, with the reputation range fixed at $[0,100]$. Changing $A$ modifies the proportions of agents assigned to the low- and high-reputation categories and therefore shifts the stationary cooperation level quantitatively. Nevertheless, positive $\eta$ promotes cooperation in every panel, and the cooperation-promoting effect becomes stronger as $\delta$ increases. The principal conclusions are therefore robust to the choice of the reputation threshold.

The threshold determines how frequently the two branches of the asymmetric reputation rule are applied, but it does not directly enter the adaptive exploration rule. Moreover, as shown by the local fitness comparison in Appendix~\ref{app:checkerboard}, $A$ does not appear explicitly in the one-step fitness difference in Eq.~(\ref{eq:appendix_diff}). Changing $A$ therefore affects the occupancy of the two reputation categories without altering the basic directional incentive generated by asymmetric reputation updating.

\section{Sensitivity to Learning Parameters} \label{app:learning_parameters}

\begin{figure*}[htbp]
\centering
\includegraphics[width=1\textwidth]{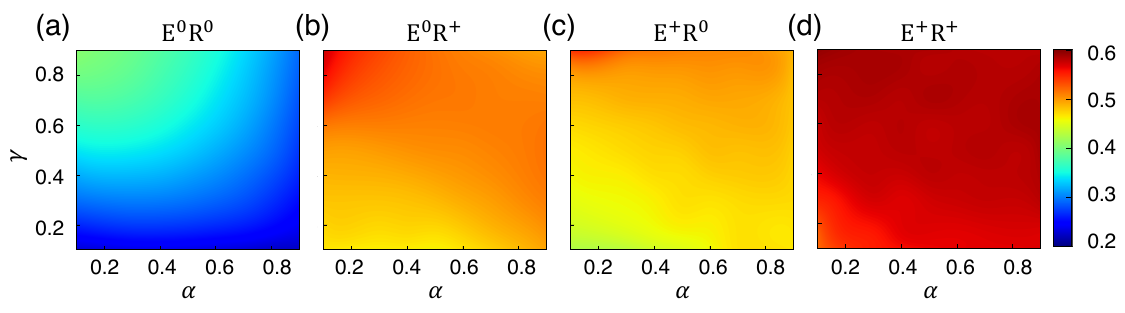}
\caption{Heat maps of the fraction of cooperation $\rho_{\mathrm{C}}$ in the $(\alpha,\gamma)$ plane for four mechanism combinations. A smaller learning rate $\alpha$ and a larger discount factor $\gamma$ generally favor cooperation. Individual mechanisms are relatively sensitive to learning parameters, but when combined ($\mathrm{E}^{+}\mathrm{R}^{+}$), cooperation is maintained at a high level throughout the parameter space, confirming the robustness of the cooperative advantage of the combined mechanism. The fixed parameters are $b=1.6$, $\theta=0.6$, $\epsilon_0=0.02$.}
\label{fig:learn}
\end{figure*}

\begin{figure*}[htbp]
\centering
\includegraphics[width=0.9\textwidth]{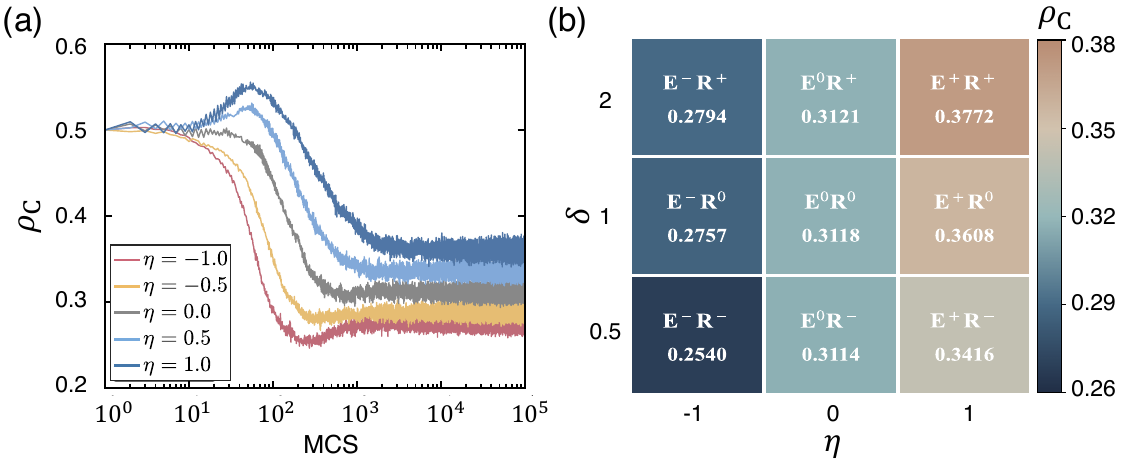}
\caption{Mechanism isolation at $\theta=0$. (a) Time evolution of the fraction of cooperation $\rho_{\mathrm{C}}$ for different exploration biases $\eta$ under symmetric reputation updating ($\delta=1$). When $\eta > 0$, cooperation is promoted. When $\eta < 0$, cooperation is inhibited. (b) Nine combinations of exploration mechanism and reputation update rule, where each cell reports the corresponding $\rho_{\mathrm{C}}$. Even without reputation's direct fitness effect, the reinforcing relationship persists, as $\rho_{\mathrm{C}}$ under $\mathrm{E}^{+}\mathrm{R}^{+}$ exceeds both $\mathrm{E}^{+}\mathrm{R}^{0}$ and $\mathrm{E}^{0}\mathrm{R}^{+}$. The fixed parameters are $b=1.6$, $\epsilon_0=0.02$.}
\label{fig:theta0}
\end{figure*}

To examine how learning parameters affect the evolution of cooperation, we scan the two-dimensional parameter space of learning rate $\alpha\in[0.1,0.9]$ and discount factor $\gamma\in[0.1,0.9]$. The learning rate $\alpha$ governs the responsiveness of Q-value updates, while the discount factor $\gamma$ captures how much agents value future rewards.

Figure~\ref{fig:learn} presents heat maps of the cooperation frequency for four mechanism combinations. All four panels reveal a consistent pattern in which a smaller learning rate and a larger discount factor generally favor cooperation~\cite{zheng2026brief}. A low learning rate makes Q-value updates more cautious, preventing agents from prematurely locking into defection strategies. A high discount factor encourages agents to value long-term payoffs, allowing the reputation accumulation and future reciprocity generated by cooperation to be fully realized.

A comparison across the four panels indicates that individual mechanisms ($\mathrm{E}^{+}\mathrm{R}^{0}$ or $\mathrm{E}^{0}\mathrm{R}^{+}$) are relatively sensitive to learning parameters. However, when the two mechanisms are combined ($\mathrm{E}^{+}\mathrm{R}^{+}$), cooperation is maintained at a high level throughout the parameter space. This demonstrates that the synergistic effect persists across a wide range of $\alpha$ and $\gamma$ values, confirming the robustness of our conclusions. Even as learning parameters vary, the joint action of adaptive exploration and asymmetric reputation updating consistently promotes cooperation.

\section{Mechanism Isolation without Reputation in Fitness} \label{app:mechanism_isolation}

To isolate the exploration channel from the direct contribution of reputation to fitness, we set $\theta=0$. Under this condition, reputation does not enter the learning reward, but it can still regulate action exploration through Eq.~(\ref{eq:exploration}).

Figure~\ref{fig:theta0}(a) shows the evolution of cooperation for different exploration biases under symmetric reputation updating ($\delta=1$). Even when reputation serves only as an exploration-modulating signal, $\eta>0$ produces a higher stationary cooperation level than the fixed-exploration baseline $\eta=0$, whereas $\eta<0$ suppresses cooperation. This confirms that the directional effect of reputation-based adaptive exploration does not require a direct contribution of reputation to fitness.

Figure~\ref{fig:theta0}(b) further reveals the interplay between exploration mechanisms and reputation updating rules at $\theta=0$. When adaptive exploration is combined with asymmetric updating ($\mathrm{E}^{+}\mathrm{R}^{+}$), the cooperative enhancement exceeds the effect produced by either mechanism in isolation. This indicates that the reinforcing relationship between adaptive exploration and asymmetric reputation updating persists even when reputation does not contribute directly to fitness.

\section{Checkerboard-Like Pattern Analysis}\label{app:checkerboard}

\subsection{Local Fitness Comparison}

We first provide a local fitness comparison that helps explain the emergence and stability of the checkerboard-like coexistence shown in Fig.~\ref{fig:fig5}(b).

We consider a focal agent with reputation $R$ and $n_\mathrm{C}\in\{0,1,2,3,4\}$ cooperative neighbors. Under the weak PDG, the one-step payoff is $P_\mathrm{C}=n_\mathrm{C}$ if the agent cooperates and $P_\mathrm{D}=bn_\mathrm{C}$ if it defects. The fitness is given by Eq.~(\ref{eq:fitness}), where the reputation term uses the post-update reputation.

According to the reputation rule in Eq.~(\ref{eq:reputation}), the reputation change depends on the current status. If $R<A$, cooperation yields $R' = R+\delta$ while defection yields $R' = R-1$. If $R\ge A$, cooperation yields $R' = R+1$ while defection yields $R' = R-\delta$. In both cases, the difference between choosing cooperation and defection is the same,
\begin{equation}\label{eq:rdiff}
R'_\mathrm{C} - R'_\mathrm{D} = \delta + 1.
\end{equation}
Using Eq.~(\ref{eq:fitness}), the one-step fitness difference between cooperation and defection can be written as
\begin{equation}
\begin{aligned}
    f_\mathrm{C} - f_\mathrm{D}
    &= (1-\theta)(n_\mathrm{C} - bn_\mathrm{C}) + \theta \frac{4b}{R_{\max}-R_{\min}}(R'_\mathrm{C} - R'_\mathrm{D})\\
    &= \theta \frac{4b}{R_{\max}-R_{\min}}(\delta+1) - (1-\theta)n_\mathrm{C}(b-1).  
\end{aligned} 
\label{eq:appendix_diff}
\end{equation}
This expression implies a critical neighbor count
\begin{equation}
n_\mathrm{C}^{\ast}
= \frac{\theta}{1-\theta}\frac{4b(\delta+1)}{(R_{\max}-R_{\min})(b-1)},
\label{eq:appendix_ncstar}
\end{equation}
such that cooperation is favored when $n_\mathrm{C}<n_\mathrm{C}^{\ast}$ and defection is favored when $n_\mathrm{C}>n_\mathrm{C}^{\ast}$.

A checkerboard-like coexistence requires that cooperation is advantageous in defector-rich surroundings, while defection can still be advantageous in cooperator-rich surroundings. A sufficient condition is $0<n_\mathrm{C}^{\ast}<4$. For Fig.~\ref{fig:fig5}(b) with $\theta=0.6$, $\delta=3$, $b=1.5$, and $R_{\max}-R_{\min}=100$, Eq.~(\ref{eq:appendix_ncstar}) gives $n_\mathrm{C}^{\ast}=0.72$. This yields $f_\mathrm{C}-f_\mathrm{D}>0$ at $n_\mathrm{C}=0$ and $f_\mathrm{C}-f_\mathrm{D}<0$ at $n_\mathrm{C}=4$, which supports an alternating arrangement. For Fig.~\ref{fig:fig5}(c) with $\theta=0.9$ and the same $\delta$, $b$, and reputation range, Eq.~(\ref{eq:appendix_ncstar}) gives $n_\mathrm{C}^{\ast}=4.32>4$, so $f_\mathrm{C}-f_\mathrm{D}$ remains nonnegative for all $n_\mathrm{C}\in\{0,1,2,3,4\}$. In this case, the alternating pattern is not stable and the system tends toward near-full cooperation.

Finally, an ideal checkerboard would give $\rho_\mathrm{C}\simeq 0.5$, whereas our simulations show a checkerboard-like state with $\rho_\mathrm{C}\simeq 0.6$. This deviation is consistent with the joint effect of adaptive exploration and asymmetric reputation updating. Low-reputation agents explore more frequently, so defectors embedded in the coexistence structure more often try cooperation. Under $\delta>1$, successful cooperative trials yield faster reputation recovery, which reduces subsequent exploration and makes cooperation more persistent. As a result, some sites that would be defectors in an ideal alternating configuration become cooperators, producing a cooperator-enriched checkerboard-like pattern.

\subsection{Spatiotemporal Analysis of the Checkerboard Structure}

To further characterize the checkerboard structure, we adopt the spatial configuration analysis framework developed by Li~{\it et~al}\cite{li2025cooperation}.

To characterize the behavioral tendencies of agents in the checkerboard-like structure, we first define the time-averaged cooperation tendency of agent $i$ during the stationary stage:
\begin{equation}
\rho_\mathrm{C}^i = \frac{1}{T}\sum_{\tau=t_0}^{t_0+T-1} s_i^\mathrm{C}(\tau),
\end{equation}
where $s_i^\mathrm{C}(\tau)=1$ indicates that agent $i$ chooses cooperation at time step $\tau$, and $0$ otherwise. Here, $t_0$ marks the end of the transient period (taken as the first $9.5\times10^{4}$ MCS), and $T$ is the duration of the stationary statistical window (the last $5\times10^{3}$ MCS). Based on $\rho_\mathrm{C}^i$, we partition the population into two groups: the inclined cooperators set $\mathcal{N}_\mathrm{C} = \{i\mid \rho_\mathrm{C}^i \geq 0.5\}$, whose behavior is predominantly cooperative over long timescales, and the inclined defectors set $\mathcal{N}_\mathrm{D} = \{i\mid \rho_\mathrm{C}^i < 0.5\}$, whose behavior is predominantly defective.

Building upon the above classification, we examine the state distributions of $\mathcal{N}_\mathrm{C}$ and $\mathcal{N}_\mathrm{D}$. For each group, the time-averaged proportion of agents in state $s$ during the stationary stage is calculated as:
\begin{align}
\bar{f}_s^\mathrm{C} &:= \frac{\sum_{\tau=t_0}^{t_0+T-1} \sum_{i\in\mathcal{N}_\mathrm{C}} \mathbf{1}_{s_i(\tau)=s}}{|\mathcal{N}_\mathrm{C}|T},\\
\bar{f}_s^\mathrm{D} &:= \frac{\sum_{\tau=t_0}^{t_0+T-1} \sum_{i\in\mathcal{N}_\mathrm{D}} \mathbf{1}_{s_i(\tau)=s}}{|\mathcal{N}_\mathrm{D}|T}.
\end{align}
Here, $\mathbf{1}_{\mathrm{predicate}}$ denotes the random variable that is $1$ if predicate is true and $0$ if it is not, where $s\in\{\text{HC},\text{LC},\text{HD},\text{LD}\}$.

\begin{figure*}[htbp]
\centering
\includegraphics[width=1\textwidth]{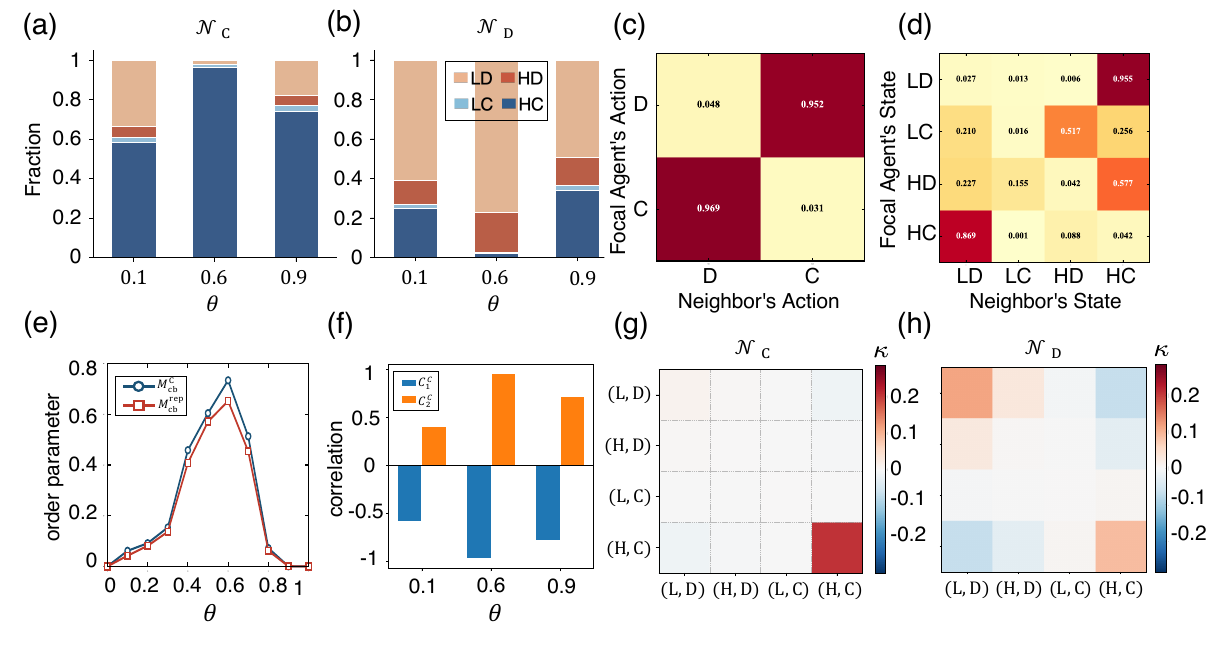}
\caption{State composition, spatial structure, and temporal persistence of the checkerboard-like coexistence. (a)--(b) Fractions of the four states in $\mathcal{N}_\mathrm{C}$,$\mathcal{N}_\mathrm{D}$ versus $\theta$. When $\theta=0.6$, $\mathcal{N}_\mathrm{C}$ is dominated by high-reputation cooperators (HC) and $\mathcal{N}_\mathrm{D}$ by low-reputation defectors (LD), consistent with the checkerboard-like pattern in Fig.~\ref{fig:fig5}(b). (c)--(d) Nearest-neighbor anti-coordination matrix $P_{ss'}^{\mathrm{nn}}$ for the two actions, and four states. The dominant off-diagonal entries confirm that HC and LD states are preferentially arranged as nearest neighbors. (e) Checkerboard order parameters $M_{\mathrm{cb}}^\mathrm{C}$ and $M_{\mathrm{cb}}^{\mathrm{rep}}$ as functions of $\theta$. Both quantities peak at intermediate $\theta$, identifying the parameter regime in which checkerboard-like spatial ordering is strongest, consistent with the spatiotemporal snapshots in Fig.~\ref{fig:fig5}. (f) Nearest-neighbor spatial correlation $C_1^C$ and next-nearest-neighbor correlation $C_2^\mathrm{C}$ as functions of $\theta$. At $\theta=0.6$, both $|C_1^\mathrm{C}|$ and $|C_2^\mathrm{C}|$ are close to 1, confirming that nearest neighbors are strongly anti-correlated while next-nearest neighbors are positively correlated, consistent with the checkerboard-like anti-coordination. (g)--(h) Cohen's $\kappa$ matrices for $\mathcal{N}_\mathrm{C}$ and $\mathcal{N}_\mathrm{D}$ at $\theta=0.6$, showing positive diagonal elements that indicate temporal self-loop persistence rather than rapid HC--LD switching. The fixed parameters are $b=1.5$, $\delta=3$, $\eta=1$, and $\epsilon_0=0.02$.}
\label{fig:appendix_chessboard}
\end{figure*}

Figures~\ref{fig:appendix_chessboard}(a)--(b) present the state distributions for three representative values of $\theta$. When $\theta=0.6$, the state compositions of $\mathcal{N}_\mathrm{C}$ and $\mathcal{N}_\mathrm{D}$ exhibit HC-LD differentiation: $\mathcal{N}_\mathrm{C}$ is almost entirely dominated by high-reputation cooperators (HC), whereas $\mathcal{N}_\mathrm{D}$ primarily consists of low-reputation defectors (LD). This HC-LD differentiation is consistent with the checkerboard-like pattern observed in Fig.~\ref{fig:fig5}(b).

However, state distributions alone can only demonstrate that HC and LD are the dominant types in $\mathcal{N}_\mathrm{C}$ and $\mathcal{N}_\mathrm{D}$, respectively. They cannot establish that these two types are spatially interleaved in a checkerboard arrangement. To verify the checkerboard structure, we therefore need spatial correlation metrics that directly probe nearest-neighbor relationships.

\subsection{Nearest-neighbor Anti-coordination Matrix}

We define a nearest-neighbor anti-coordination matrix to quantify the spatial arrangement between different states. For the four-state case with $s,s'\in\{\text{HC},\text{LC},\text{HD},\text{LD}\}$, the time-averaged nearest-neighbor conditional probability is:
\begin{equation}
P_{ss'}^{\mathrm{nn}}
=
\frac{\displaystyle\sum_{\tau=t_0}^{t_0+T-1} \sum_{i}\sum_{j\in\Omega_i}
      \mathbf{1}_{x_i(\tau)=s}\,
      \mathbf{1}_{x_j(\tau)=s'}}
     {\displaystyle\sum_{\tau=t_0}^{t_0+T-1} \sum_{i}\sum_{j\in\Omega_i}
      \mathbf{1}_{x_i(\tau)=s}},
\label{eq:nn_matrix}
\end{equation}
where $\mathbf{1}_{x_i(\tau)=s}$ is the indicator that agent $i$ is in state
$s$ at time step $\tau$, $\Omega_i$ denotes the set of nearest neighbors of
$i$. The resulting $P_{ss'}^{\mathrm{nn}}$ is the time-averaged
nearest-neighbor conditional probability between states $s$ and $s'$.

Figures~\ref{fig:appendix_chessboard}(c)--(d) show the nearest-neighbor anti-coordination matrix at $\theta=0.6$. The dominant entries are $P_{CD}^{\mathrm{nn}} = 0.969$, $P_{DC}^{\mathrm{nn}} = 0.952$ and $P_{\mathrm{HC},\mathrm{LD}}^{\mathrm{nn}} = 0.869$, $P_{\mathrm{LD},\mathrm{HC}}^{\mathrm{nn}} = 0.955$. This pattern confirms that HC and LD states are preferentially arranged as nearest neighbors rather than forming large same-type clusters. It provides direct spatial evidence for the anti-coordination between high-reputation cooperators and low-reputation defectors in the checkerboard-like structure.

\subsection{Checkerboard Order Parameters}

To quantify how the strength of checkerboard-like spatial ordering varies with $\theta$, we define two antiferromagnetic order parameters.

The first is based on the cooperation tendency $\rho_\mathrm{C}^i$:
\begin{equation}
M_{\mathrm{cb}}^\mathrm{C} = \left|\frac{1}{N}\sum_i (2\rho_\mathrm{C}^i - 1)(-1)^{x_i+y_i}\right|,
\label{eq:mcb_c}
\end{equation}
where $(x_i,y_i)$ are the lattice coordinates of agent $i$. If inclined cooperators and defectors alternate on the lattice, the product $(2\rho_\mathrm{C}^i-1)(-1)^{x_i+y_i}$ has the same sign across sites, yielding a large $M_{\mathrm{cb}}^\mathrm{C}$. If the population is uniformly cooperative, uniformly defective, or randomly mixed, the sum averages to a small value.

Since the checkerboard in our model involves not only cooperation and defection but also high and low reputation, we define a second order parameter that captures the HC-LD anti-coordination directly:
\begin{equation}
M_{\mathrm{cb}}^{\mathrm{rep}} = \left|\frac{1}{N}\sum_i (\bar{\mathbf{1}}_{i,\mathrm{HC}} - \bar{\mathbf{1}}_{i,\mathrm{LD}})(-1)^{x_i+y_i}\right|,
\label{eq:mcb_rep}
\end{equation}
where $\bar{\mathbf{1}}_{i,\mathrm{HC}} = \frac{1}{T}\sum_{\tau=t_0}^{t_0+T-1}\mathbf{1}_{x_i(\tau)=\mathrm{HC}}$ and $\bar{\mathbf{1}}_{i,\mathrm{LD}} = \frac{1}{T}\sum_{\tau=t_0}^{t_0+T-1}\mathbf{1}_{x_i(\tau)=\mathrm{LD}}$ are the time-averaged indicators for the HC and LD states, respectively.

Figure~\ref{fig:appendix_chessboard}(e) shows $M_{\mathrm{cb}}^{\mathrm{C}}$, $M_{\mathrm{cb}}^{\mathrm{rep}}$ as functions of $\theta$. At low $\theta$, $M_{\mathrm{cb}}^\mathrm{C}$ and $M_{\mathrm{cb}}^{\mathrm{rep}}$ are small, reflecting the prevalence of defection with no spatial ordering. As $\theta$ increases toward intermediate values, both order parameters rise, indicating the emergence of the checkerboard-like anti-coordination. At large $\theta$, both order parameters decline as the population shifts toward near-full cooperation. This non-monotonic behavior identifies an intermediate range of $\theta$ over which checkerboard-like ordering is most pronounced. At smaller $\theta$, reputation is too weak to sustain HC--LD spatial differentiation, whereas at larger $\theta$, the expansion of cooperation suppresses the alternating structure. This result is consistent with the spatiotemporal snapshots in Fig.~\ref{fig:fig5}.

\subsection{Nearest-neighbor Spatial Correlation}

To further characterize the spatial structure, we define a nearest-neighbor spatial correlation function. Let $\bar{\rho}_\mathrm{C} = \frac{1}{N}\sum_i \rho_\mathrm{C}^i$ be the global mean cooperation tendency. The nearest-neighbor correlation is:
\begin{equation}
C_1^\mathrm{C} = \frac{\left\langle (\rho_\mathrm{C}^i - \bar{\rho}_\mathrm{C})(\rho_\mathrm{C}^j - \bar{\rho}_\mathrm{C})\right\rangle_{\langle i,j\rangle}}{\left\langle (\rho_\mathrm{C}^i - \bar{\rho}_\mathrm{C})^2\right\rangle_i},
\label{eq:c1}
\end{equation}
where $\langle i,j\rangle$ denotes averaging over all nearest-neighbor pairs. A checkerboard-like anti-coordination implies that neighboring agents tend to have opposite signs of $\rho_\mathrm{C}^i - \bar{\rho}_\mathrm{C}$, so one expects $C_1^\mathrm{C} < 0$.

Similarly, we define the next-nearest-neighbor correlation:
\begin{equation}
C_2^\mathrm{C} = \frac{\left\langle (\rho_\mathrm{C}^i - \bar{\rho}_\mathrm{C})(\rho_\mathrm{C}^j - \bar{\rho}_\mathrm{C})\right\rangle_{\langle\langle i,j\rangle\rangle}}{\left\langle (\rho_\mathrm{C}^i - \bar{\rho}_\mathrm{C})^2\right\rangle_i},
\label{eq:c2}
\end{equation}
where $\langle\langle i,j\rangle\rangle$ denotes averaging over next-nearest-neighbor pairs (diagonal neighbors on the square lattice). In a checkerboard-like arrangement, next-nearest neighbors tend to be of the same type, so $C_2^\mathrm{C} > 0$.

Figure~\ref{fig:appendix_chessboard}(f) shows $C_1^\mathrm{C}$ and $C_2^\mathrm{C}$ for three representative values of $\theta$. At $\theta=0.6$, both $|C_1^\mathrm{C}|$ and $|C_2^\mathrm{C}|$ are close to 1, confirming that nearest neighbors are strongly anti-correlated while next-nearest neighbors are positively correlated, consistent with the checkerboard-like anti-coordination. At $\theta=0.1$ and $\theta=0.9$, $|C_1^\mathrm{C}|$ and $|C_2^\mathrm{C}|$ are smaller, reflecting weaker spatial ordering.

\subsection{Temporal Persistence of the Checkerboard Structure}

The above spatial metrics establish the existence of the checkerboard pattern. We now turn to its temporal persistence. To this end, we define the Cohen's $\kappa$ coefficient for state transitions:
\begin{equation}
\kappa(s,s') = \frac{\bar{f}(s,s') - \bar{f}(s)\,\bar{f}(s')}{1 - \bar{f}(s)\,\bar{f}(s')},
\end{equation}
where $\bar{f}(s)$ and $\bar{f}(s,s')$ denote the marginal frequency of state $s$ and the joint frequency of consecutive time steps $(s,s')$, respectively. Diagonal elements $\kappa(s,s)>0$ indicate temporal self-loop persistence: an agent in state $s$ tends to remain in state $s$ at the next time step. Symmetric off-diagonal elements $\kappa(s,s')>0$ and $\kappa(s',s)>0$ indicate period-2 oscillatory cycles between states $s$ and $s'$.

Figures~\ref{fig:appendix_chessboard}(g)--(h) display the $\kappa$ matrices for $\mathcal{N}_\mathrm{C}$ and $\mathcal{N}_\mathrm{D}$ at $\theta=0.6$. The positive diagonal elements for $(H,C)\leftrightarrow(H,C)$ in $\mathcal{N}_\mathrm{C}$ and $(L,D)\leftrightarrow(L,D)$ in $\mathcal{N}_\mathrm{D}$ are most prominent, indicating temporal self-loop persistence rather than period-2 oscillations. This means that the checkerboard-like structure is not maintained by rapid HC–LD switching at individual sites, but by the persistence of two spatially anti-coordinated state identities (the HC-dominated inclined cooperators and the LD-dominated inclined defectors).

In summary, the checkerboard-like coexistence is established by the nearest-neighbor anti-coordination matrices (Figs.~\ref{fig:appendix_chessboard}(c)--(d)), quantified by the checkerboard order parameters (Fig.~\ref{fig:appendix_chessboard}(e)) and spatial correlation function (Fig.~\ref{fig:appendix_chessboard}(f)), and explained temporally by the self-loop persistence captured by Cohen's $\kappa$ (Figs.~\ref{fig:appendix_chessboard}(g)--(h)). Together, these results identify an intermediate reputation-concern regime in which HC--LD spatial anti-coordination is both pronounced and temporally persistent.

% \bibliographystyle{}
% \bibliography{ref}

\end{document}